\documentclass[twocolumn,journal]{IEEEtran}
\usepackage[latin9]{inputenc}
\usepackage{amsmath}
\usepackage{amssymb}
\usepackage{graphicx}
\usepackage[unicode=true]
 {hyperref}
\usepackage{breakurl}

\usepackage{fancyhdr}
\makeatletter
\IEEEoverridecommandlockouts
\usepackage{cite}
\usepackage{amsfonts}\usepackage{algorithmic}
\usepackage{subfigure}
\usepackage{textcomp}
\usepackage{lipsum}
\usepackage{mathtools}
\usepackage{cuted}
\usepackage{xcolor}
\def\BibTeX{{\rm B\kern-.05em{\sc i\kern-.025em b}\kern-.08em
    T\kern-.1667em\lower.7ex\hbox{E}\kern-.125emX}}
\newlength\wttlinewidth 
\newlength\parindentlength
\makeatother

\begin{document}

\title{Analytical Solution for Space-Charge Waves in a Two-Stream Cylindrical
Electron Beam}

\author{\IEEEauthorblockN{Tarek~Mealy, Robert~Marosi, Kasra~Rouhi, and Filippo\ Capolino}
\thanks{Tarek\ Mealy, Robert~Marosi, Kasra~Rouhi, and Filippo\ Capolino
are with the Department of Electrical Engineering and Computer Science,
University of California, Irvine, Irvine, California, e-mail: \protect\href{mailto:tmealy@uci.edu}{tmealy@uci.edu}~\protect\href{mailto:rmarosi@uci.edu}{rmarosi@uci.edu}\ \protect\href{mailto:kasra.rouhi@uci.edu}{kasra.rouhi@uci.edu}\ \protect\href{mailto:f.capolino@uci.edu}{f.capolino@uci.edu}.}}
\maketitle

\thispagestyle{fancy}

\begin{abstract}
\noindent We present an analytical method to compute the wavenumbers
and electric fields of the space-charge-wave eigenmodes supported
by a two-stream electron beam, consisting of a solid inner cylindrical
stream and a coaxial outer annular stream, both contained within a
cylindrical metallic tunnel. We extend the analytical model developed
by Ramo to the case of two streams. The method accounts for the interaction
between the two streams with the presence of the beam-tunnel wall;
it can be used to model the complex wavenumbers associated with the
two-stream instability and the plasma frequency reduction effects
in vacuum electronic amplifiers and other vacuum electronic devices.
\end{abstract}

\begin{IEEEkeywords}
Bifurcation points; Double-stream; Eigenmode solution; Electron beam
devices; Exceptional points of degeneracy; Transition points; Two-stream
instability.

{\let\thefootnote\relax\footnotetext{This material is based upon work supported by the Air Force Office of Scientific Research award number FA9550-18-1-0355 and by the MURI Award number FA9550- 20-1-0409 administered through the University of New Mexico.}} 
\end{IEEEkeywords}

\vphantom{}

\section{Introduction}

\IEEEPARstart{V}{acuum}electron devices with high power and broad
bandwidth have a competitive edge in various applications, such as
electronic countermeasures, satellite communication, plasma diagnostics,
and high-resolution radars \cite{qiu2009vacuum,booske2011vacuum}.
Lately, the designers of microwave tubes have faced many difficult
design challenges, such as reducing operating voltages and minimizing
the weight and dimensions of the devices and their power supplies.
In addition, with the high demand for vacuum electronics applications
that operate at high frequencies, the dimensions of these devices
are being reduced, and at the same time, electron beams with high
current density are also required to obtain high output power \cite{booske2008plasma}.
There are some technical limitations to increasing both the output
power and the operating frequency. The product $Pf^{2}$ ($P$ is
the average output power and $f$ is the frequency of operation),
also called the ``power density'' of the state-of-the-art vacuum
electronics is a figure of merit that tends to follow a growing linear
trend with time \cite{parker2002vacuum}. However, it is not certain
how long vacuum electronic devices will continue to follow this trend.
A promising engineering solution to continue improving the power density
of vacuum electronic devices is to use multiple electron beams \cite{pobedonostev1993multiple,sheng1998multi}.

The interaction of multi-stream electron beams has been studied theoretically
for years. Many authors have also proposed the multiple beam concept
since the 1940s for use in electron beam devices. As a pioneer in
this field, Pierce predicted the gain of a double-stream amplifier
having thin concentric electron streams of different velocities that
are modulated by input and output cavities. \cite{pierce1949double}.
Then, Swift-Hook analyzed the validity of the theory of the double
stream amplification model proposed by Pierce \cite{swift1960validity}.
As an early work on this topic, beam-beam interaction in concentric-beam
dual-mode traveling-wave tubes (TWTs) is presented in \cite{dohler1980beam}
and then investigated in more detail for various kinds of TWTs in
\cite{sheng1998multi}. Chen analyzed the conversion mechanism from
the kinetic energy of electron beams to electromagnetic wave energy
in the two-stream amplifier and how the efficiency of a two-stream
instability amplifier increases with relativistic beam velocities
\cite{chen1996efficiency}. Wave coupling in multiple beam TWTs to
increase the power level of vacuum electronic devices has also been
studied in \cite{nusinovich2009wave}. On the other hand, many works
have begun exploring and showing realistic structures for multi-beam
generation. In \cite{zavadil1974dual}, Zavadil proposed a dual-cathode
electron gun incorporating an annular hollow beam cathode, concentric
and co-planar with a solid beam cathode. Some work has used multiple
cathode sources to produce multiple electron beams in low-power microwave
sources where two separate power supplies power each cathode at different
voltages \cite{hollenberg1949experimental,neben2021co}. Also, some
work has been published in the past that uses conventional vacuum
electron beam device concepts to generate multiple electron beams
\cite{nergaard1948analysis,haeff1949electron,butler1996twin,carlsten2008compact}.
Multiple electron beam generation with comparable currents and different
energies from a single cathode-anode voltage for high power applications
has been studied recently in \cite{islam2022multiple,islam2022modeling}.
Another significant motivation for our work has been the analytic
theory developed for multi-stream electron beam devices in \cite{figotin2021analytic,evans2019instability},
and the theoretical work involving modal degeneracies in linear beam
tubes \cite{abdelshafy2018electron,figotin2013multi,figotin2021exceptional,othman2016low,rouhi2021exceptional,yazdi2017new}.

Modern communications' increasing range and data handling requirements
have given rise to a need for microwave tubes with power output and
bandwidth capabilities that greatly exceed those of present-day state-of-the-art
single-stream electron beam devices. The multiple-beam concept was
developed to address this need and was applied to a resonant klystron
\cite{boyd1962multiple}, which was demonstrated to be capable of
an order of magnitude higher power output than single-beam devices
using the same electron beam. Then the development of multi-beam klystron
to provide low operating voltages, high power, low noise, and the
possibility of larger operating bandwidth is further studied in many
papers such as \cite{pohl1965design,pobedonostev1993multiple,ding2005s}.
Recently, several research papers have focused on electron beam devices
that utilize multiple electron streams, namely, multi-beam folded
waveguide structures \cite{yan2014design,yan2017design,liu2018nonlinear,shi2019study},
two-stream gyrotron TWT amplifiers \cite{yang1999two}, staggered
dual-beam waveguides \cite{shin2008strongly,gee2013gain,yang2019staggered,shao2019stacked,lu2020novel},
dual-beam sine waveguide TWTs \cite{lu2020investigation,luo2021340},
and other unique TWT configurations \cite{wang2018study,torgashov2019meander,wen2021concentric,liao2022terahertz,torgashov2022design}.
The aforementioned devices can employ the advantages of two-stream
beams to improve the output power significantly and/or increase the
bandwidth for various applications such as telecommunication and high-resolution
radar.

\begin{figure*}[tbh]
\begin{centering}
\centering \subfigure[]{\includegraphics[width=0.62\textwidth]{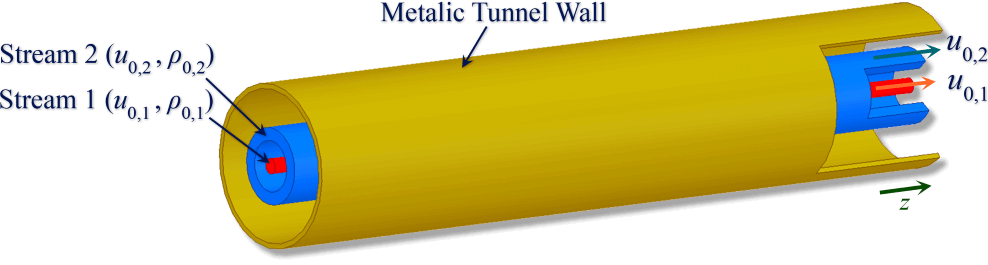}\label{Fig:General_Setup_a}}\subfigure[]{\includegraphics[width=0.38\textwidth]{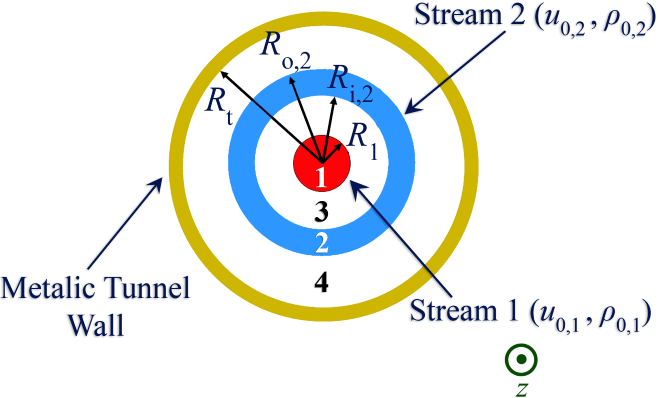}\label{Fig:Front_View}}
\par\end{centering}
\centering{}\caption{(a) Double stream electron beam with distinct dc speed and charge
density. We show in (b) the transverse cross section for the double
stream system, where the area inside of the tunnel is divided into
four homogeneous regions. Stream 1 (the inner stream) exists in region
1, whereas stream 2 (the outer stream) exists in region 2, and regions
3 and 4 are vacuum spaces.\label{fig:Schematic}}
\end{figure*}

The problem of space-charge waves in an electron beam is a topic of
interest since such waves are excited and utilized in a variety of
electron tubes. These tubes may be used to generate, amplify, and
detect signals. Furthermore, such tubes utilizing multi-stream electron
beams may either be designed to utilize or avoid strong coupling between
electron streams that leads to the two-stream instability under certain
conditions. Two-stream instability conditions depend on the velocity
differences between electron streams, their respective current densities,
operating frequency, and geometry \cite{pierce1949double}. In particular,
Pierce uses an \emph{ad-hoc} separation parameter $S$ to model how
strongly electron streams are coupled when they are close together
in a multi-stream beam, which affects the growth rate of space-charge
waves under the two-stream instability regime \cite{pierce1949double}.
However, no simple models have yet been developed to analytically
determine the growth rate and conditions for two-stream instability
as a function of stream geometry, voltage, and current. Ramo studied
the propagation of space-charge waves for the case of a single electron
beam propagating within a metallic beam tunnel \cite{ramo1939space}.
The theory of Ramo was extended in \cite{branch1955plasma}, where
they defined the plasma frequency reduction factor and considered
the case of an annular electron beam within a cylindrical metallic
tunnel. Here, we extend the work of Ramo to the case of two concentric
electron streams within a metallic tunnel, which allows us to analytically
determine the conditions for two-stream instability and its growth
rates without using \emph{ad-hoc} parameters. We consider an electron
beam composed of a solid stream inside a hollow coaxial stream as
in Fig. \ref{fig:Schematic}. Knowledge of the complex propagation
constants of space-charge waves supported by the two-stream system
may be useful for designing and analyzing two-stream instability amplifiers
and two-stream TWTs, which depend strongly on the geometric configuration
of the two-stream electron beam.

\section{Formulation}

\subsection{Problem Setup}

The electron beam is assumed to be made of two concentric streams:
stream 1 (the inner stream) is solid with a circular cross section
and it exists for $0\leq r\leq R_{1}$; stream 2 (the outer stream,
coaxial with stream 1) has an annular cross section and it exists
for $R_{\mathrm{i},2}\leq r\leq R_{\mathrm{o},2}$. The beam tunnel
is assumed to be cylindrical with radius $R_{\mathrm{t}}$ and made
of a perfect electric conductor (PEC), as shown in Fig. \ref{fig:Schematic}.
For convenience, we use a cylindrical coordinate system $(r,\theta,z)$
in this paper to represent both the beam and the electromagnetic fields.

The two streams are assumed to possess uniform dc charge densities
of $\rho_{0,1}$ and $\rho_{0,2}$ in both transverse and longitudinal
directions. The uniform axial dc magnetic field is assumed to be strong
enough to confine each of the two streams such that all charges travel
in the axial direction only (a common simplifying assumption seen
in other linear beam tube work such as \cite{ramo1939space,pierce1951waves,pierce1949double,pierce1947theoryTWT,chu1948field})
with dc velocities $u_{0,1}$ and $u_{0,2}$ for stream 1 and stream
2, respectively. Analogously, the existence of a strong axial dc magnetic
field leads also to the assumption that the ac modulation in the velocity
of electrons is only in the axial direction. Therefore, the radial
and the azimuthal components of electron velocities are assumed to
be vanishing \cite{ramo1939space,pierce1951waves,pierce1949double,pierce1947theoryTWT,chu1948field}.
It is our goal to find the eigenmodes that represent the space-charge
waves in this configuration. The propagating space charge wave consists
of a modulation in the beam's volumetric charge density and axial
velocity, as well as its associated electromagnetic fields, that,
in the phasor domain, are all proportional to the wave function $e^{j(\omega t-kz)}$.
As another simplifying assumption, we only consider modes with azimuthal
symmetry; therefore, we assume that $\partial/\partial\theta=0$ for
all beam and field quantities. However, the presented formalism could
be extended to find modes that do not possess azimuthal symmetry.
Nevertheless, the case of azimuthal symmetry is the most significant
one in practice for TWTs operation. Therefore, the instantaneous total
(both dc and ac components) axial velocity and volumetric charge density
for each stream are written as

\begin{equation}
\begin{array}{c}
u_{1}(r,z,t)=u_{0,1}+\Re\left(u_{\mathrm{m,1}}(r)e^{j\omega t-jkz}\right),\\
u_{2}(r,z,t)=u_{0,2}+\Re\left(u_{\mathrm{m},2}(r)e^{j\omega t-jkz}\right),
\end{array}
\end{equation}

\begin{equation}
\begin{array}{c}
\rho_{1}(r,z,t)=\rho_{0,1}+\Re\left(\rho_{\mathrm{m,1}}(r)e^{j\omega t-jkz}\right),\\
\rho_{2}(r,z,t)=\rho_{0,2}+\Re\left(\rho_{\mathrm{m,2}}(r)e^{j\omega t-jkz}\right),
\end{array}
\end{equation}
where $u_{\mathrm{m,1}}(r)$, $u_{\mathrm{m,2}}(r)$, $\rho_{\mathrm{m,1}}(r)$
and $\rho_{\mathrm{m,2}}(r)$ are the radial distributions of the
stream velocities and charge densities expressed in phasor domain,
the subscript '0' denotes the dc component, 'm' denotes ac modulation
component, '1' and '2' denote stream 1 and stream 2, respectively.
The total radial-dependent volumetric charge density inside the tunnel
is expressed as a piecewise function as

\begin{equation}
\rho(r,z,t)=\left\{ \begin{array}{l}
\begin{array}{ll}
\rho_{1}(r,z,t),\; & 0\leq r\leq R_{1}\end{array}\\
\rho_{2}(r,z,t),\;R_{\mathrm{i},2}\leq r\leq R_{\mathrm{o,2}}\\
0,\;\text{otherwise}
\end{array}\right..\label{eq:Rho_Pisewise}
\end{equation}

We assume that the ac modulation of each electron-beam stream is small
compared to the corresponding dc part. Therefore, under this small-signal
approximation, the electron beam streams have current densities in
the axial direction ($\mathbf{J}_{1}=J_{1}\hat{\mathbf{z}}$ and $\mathbf{J}_{2}=J_{2}\hat{\mathbf{z}}$)
in the form of

\begin{equation}
\begin{array}{c}
J_{1}(r,z,t)=u_{1}\rho_{1}\approx J_{0,1}+\Re\left(J_{\mathrm{m,1}}(r)e^{j\omega t-jkz}\right),\\
J_{2}(r,z,t)=u_{2}\rho_{2}\approx J_{0,2}+\Re\left(J_{\mathrm{m,2}}(r)e^{j\omega t-jkz}\right),
\end{array}
\end{equation}
where the dc current densities of stream 1 and stream 2 are $J_{0,1}=\rho_{0,1}u_{0,1}$
and $J_{0,2}=\rho_{0,2}u_{0,2}$, and the linearized ac current densities
of stream 1 and stream 2 are $J_{\mathrm{m,1}}(r)=\rho_{0,1}u_{\mathrm{m,1}}(r)+u_{0,1}\rho_{\mathrm{m,1}}(r)$
and $J_{\mathrm{m,2}}(r)=\rho_{0,2}u_{\mathrm{m,2}}(r)+u_{0,2}\rho_{\mathrm{m,2}}(r)$,
respectively. The total currents for stream 1 and stream 2 are found
using integration over each of the transverse cross sections of the
stream regions as $i_{1}(z,t)=\iint_{A_{1}}J_{1}(r,z,t)\,dA$ and
$i_{2}(z,t)=\iint_{A_{2}}J_{2}(r,z,t)\,dA$, where $A_{1}=\pi R_{1}^{2}$
and $A_{2}=\pi\left(R_{o,2}^{2}-R_{i,2}^{2}\right)$ are the cross-sectional
areas of region 1 and 2, respectively. This integration yields

\begin{equation}
\begin{array}{c}
i_{1}(z,t)=-I_{0,1}+\Re\left(I_{m,1}e^{j\omega t-jkz}\right),\\
i_{2}(z,t)=-I_{0,2}+\Re\left(I_{m,2}e^{j\omega t-jkz}\right),
\end{array}
\end{equation}
where $I_{0,1}=A_{1}\rho_{0,1}u_{0,1}$ and $I_{0,2}=A_{2}\rho_{0,2}u_{0,2}$
are the dc currents of stream 1 and stream 2, respectively, and $I_{m,1}$
and $I_{m,2}$ are the ac currents of stream 1 and stream 2 in phasor
domain, respectively.

The electromagnetic fields associated to the two-stream electron beam
are represented using the electric scalar potential and magnetic vector
potential which are expressed as

\begin{equation}
\phi(r,z,t)=\Re\left(f_{\phi}(r)e^{j\omega t-jkz}\right),\label{eq:Phi_Rep}
\end{equation}

\begin{equation}
\mathbf{A}(r,z,t)=\Re\left(f_{A}(r)e^{j\omega t-jkz}\right)\ \hat{\mathbf{z}},
\end{equation}
where $\hat{\mathbf{z}}$ is the unit vector in the $z$ direction.
The chosen magnetic vector potential has only an axial component (in
the $z$ direction) because we assume that only the longitudinal component
of current modulation is present (we neglect current directions that
are not longitudinal because we assume to have a very high, confining,
axial dc magnetic field). We use the International System of Units
(SI) in the following analysis, whereas CGS units were used in \cite{ramo1939space}.
The electric and magnetic fields in the structure are expressed in
terms of the scalar electric potential and vector magnetic potential
as $\mathbf{E}=-\nabla\phi-d\mathbf{A}/dt$ and $\mathbf{H}=\nabla\times\mathbf{A}/\mu_{0}$.
Here, we use the Lorentz gauge $\nabla\cdot\mathbf{A}=-\mu_{0}\varepsilon_{0}\partial\phi/\partial t$
\cite{balanis2012advanced_ch6}, which leads to the relation $f_{A}(r)=\dfrac{\omega\mu_{0}\varepsilon_{0}}{k}f_{\phi}(r)$,
as shown in Appendix \ref{sec:General-Solution-of-phi-stream}. Because
the tunnel region is not homogeneously filled, we represent the radially-dependent
electric scalar potential function $f_{\phi}(r)$ in Eq. (\ref{eq:Phi_Rep})
as

\begin{equation}
f_{\phi}(r)=\left\{ \begin{array}{l}
f_{\phi,1}(r),\;0\leq r\leq R_{1}\\
f_{\phi,3}(r),\;R_{\mathrm{o},1}\leq r<R_{\mathrm{i},2}\\
f_{\phi,2}(r),\;R_{\mathrm{i,2}}\leq r<R_{\mathrm{o,2}}\\
f_{\phi,4}(r),\;R_{\mathrm{o,2}}\leq r<R_{\mathrm{t}}
\end{array}\right..\label{eq:Poten_Piece_Wise}
\end{equation}

The time-domain electric and magnetic fields (which do not depend
on $\theta$ due to the assumption of azimuthal symmetry) are then
given by

\begin{equation}
\begin{array}{c}
E_{r}(r,z,t)=\Re\left(-f_{\phi}^{\prime}(r)e^{j\omega t-jkz}\right),\\
E_{z}(r,z,t)=\Re\left(\left(j\dfrac{k^{2}-\omega^{2}\mu_{0}\varepsilon_{0}}{k}\right)f_{\phi}(r)e^{j\omega t-jkz}\right),\\
H_{\theta}(r,z,t)=\Re\left(-\dfrac{\omega\varepsilon_{0}}{k}f_{\phi}^{\prime}(r)e^{j\omega t-jkz}\right),
\end{array}\label{eq:EHfields}
\end{equation}
whereas the rest of time-varying field components are vanishing, i.e.,
$E_{\theta}=H_{r}=H_{z}=0$.

\subsection{Governing Equations}

We start by writing Newton's second law, which describes the equations
of motion for each stream individually. The basic equations that govern
the charges\textquoteright{} longitudinal motion are

\begin{equation}
\begin{array}{c}
m\dfrac{du_{1}}{dt}=-eE_{z,1},\end{array}\label{eq:eq_motions_S1}
\end{equation}

\begin{equation}
m\dfrac{du_{2}}{dt}=-eE_{z,2},\label{eq:eq_motions_S2}
\end{equation}
where $E_{z,1}$ and $E_{z,2}$ are the longitudinal electric fields
that stream 1 and stream 2 experience in each of the regions denoted
by indices '1' and '2', respectively (See Fig. \ref{fig:Schematic}),
$m=9.109\times10^{-31}\:\mathrm{kg}$ is the rest mass of an electron,
and $e=+1.602\times10^{-19}\:\mathrm{C}$ is the elementary charge.
The longitudinal electric fields $E_{z,1}$ and $E_{z,2}$ are determined
from Eq. (\ref{eq:EHfields}) in region 1 and region 2, as we will
discuss later. Each electron flow should be continuous and there should
be no leakage or accumulation of charges. Therefore, the continuity
equation for each stream is written as $\nabla\cdot\mathbf{J}_{1}=-\partial\rho_{1}/\partial t$
and $\nabla\cdot\mathbf{J}_{2}=-\partial\rho_{2}/\partial t$ which
are simplified as

\begin{equation}
\dfrac{\partial\left(\rho_{1}u_{1}\right)}{\partial z}=-\dfrac{\partial\rho_{1}}{\partial t},\label{eq:New_S1}
\end{equation}

\begin{equation}
\dfrac{\partial\left(\rho_{2}u_{2}\right)}{\partial z}=-\dfrac{\partial\rho_{2}}{\partial t}.\label{eq:New_S2}
\end{equation}

As explained in Appendix \ref{sec:General-Solution-of-phi-stream},
these charge continuity and force equations lead to the velocity and
charge modulations of each stream, expressed in terms of the scalar
potential as

\begin{equation}
u_{\mathrm{m},i}(r)=\dfrac{\eta}{u_{0,i}}\dfrac{k_{0}^{2}-k^{2}}{k\left(k-\beta_{0,i}\right)}f_{\phi,i}(r),
\end{equation}

\begin{equation}
\rho_{\mathrm{m,}i}(r)=-\dfrac{\eta\rho_{0,i}}{u_{0,i}^{2}}\dfrac{k_{0}^{2}-k^{2}}{\left(k-\beta_{0,i}\right)^{2}}f_{\phi,i}(r),\label{eq:Rho_as_phi}
\end{equation}
where $i=1,2$ refers to stream 1 or 2. Furthermore, $\beta_{0,i}=\omega/u_{0,i}$
is the electronic phase constant of the $i$-th electron stream, $\eta=e/m=1.758829\times10^{11}\:\mathrm{C/kg}$
is the charge to mass ratio of an electron and $k_{0}=\omega\sqrt{\mu_{0}\varepsilon_{0}}$.

Following what was done in \cite{ramo1939space} for a single-stream
electron beam inside a concentric metallic tunnel, the governing equation
in each region of our problem is found by substituting the definition
$\mathbf{E}=-\nabla\phi-d\mathbf{A}/dt$ into Gauss' law $\nabla\cdot\mathbf{E}=\rho/\epsilon_{0}$,
and by using the Lorentz gauge (see Appendix \ref{sec:General-Solution-of-phi-stream}),
leading to 
\begin{equation}
\left(\nabla^{2}-\mu_{0}\varepsilon_{0}\frac{\partial^{2}}{\partial t^{2}}\right)\phi=-\dfrac{\rho}{\varepsilon_{0}}.\label{eq:Phi_DE}
\end{equation}

The charge density term in Eq. (\ref{eq:Phi_DE}) is either $\rho=0$,
in the two vacuum regions, or is given by Eq. (\ref{eq:Rho_as_phi})
in the two stream regions. Substituting these values for charge density
into Eq. (\ref{eq:Phi_DE}), we arrive at the Bessel differential
equations for the electric potential in both the vacuum and stream
regions, as explained further in Appendices \ref{sec:General-Solution-of-phi-stream}
and \ref{sec:General-Solution-of-phi-vacuum}. As a result, in the
two stream regions (i.e., regions 1 and 2), the potential solution
is expressed in terms of Bessel functions of the first and second
kind, and order zero. For the vacuum regions between the electron
streams and near the metallic wall (i.e., regions 3 and 4), the potential
solution is written in terms of modified Bessel functions of the first
and second kind, of order zero. An alternative formulation is based
on taking the electric field expressions in Eq. (\ref{eq:EHfields})
and the charge density expression in Eq. (\ref{eq:Rho_as_phi}) into
Gauss' law $\nabla\cdot\mathbf{E}=\rho/\epsilon_{0}$, and the Bessel
equations are determined by expressing everything in terms of $f_{\phi}(r)$.

\subsection{Boundary Conditions\label{subsec:Boundary-Conditions}}

Aside from the fact that the potential function $f_{\phi}(r)$ is
finite at $r=0$, following what was done in \cite{ramo1939space},
we also enforce that the potential function $f_{\phi}(r)$ and its
derivative are continuous across the boundaries between the concentric
regions illustrated in Fig. \ref{fig:Schematic} as

\begin{equation}
\begin{array}{c}
f_{\phi,1}(R_{1})=f_{\phi,3}(R_{1}),\ \ f_{\phi,1}^{\prime}(R_{1})=f_{\phi,3}^{\prime}(R_{1}),\\
f_{\phi,3}(R_{\mathrm{i,2}})=f_{\phi,2}(R_{\mathrm{i,2}}),\ \ f_{\phi,3}^{\prime}(R_{\mathrm{i,2}})=f_{\phi,2}^{\prime}(R_{\mathrm{i,2}}),\\
f_{\phi,2}(R_{\mathrm{o,2}})=f_{\phi,4}(R_{\mathrm{o,2}}),\ \ f_{\phi,2}^{\prime}(R_{\mathrm{o,2}})=f_{\phi,4}^{\prime}(R_{\mathrm{o,2}}).
\end{array}\label{eq:Contin}
\end{equation}

Due to the assumption that a tunnel is made of PEC, we also enforce
that the potential function vanishes at the tunnel wall

\begin{equation}
\begin{array}{c}
f_{\phi,4}(R_{\mathrm{t}})=0.\end{array}\label{eq:Wall}
\end{equation}

In the following section, we describe the solution of the potential
functions in each region. Then, we enforce the aforementioned boundary
conditions to find a linear system whose solution provides the eigenmodes
of charge waves in this system.

\section{Modal Dispersion Equation}

The general solutions of the scalar electric potential as a function
of radius in each region shown in Fig. \ref{fig:Schematic} are found
based on the derivation in Appendices \ref{sec:General-Solution-of-phi-stream}
and \ref{sec:General-Solution-of-phi-vacuum}. We write the radially-dependent
scalar potential function in Eq. (\ref{eq:Poten_Piece_Wise}) as

\begin{equation}
f_{\phi,1}(r)=c_{1}J_{0}(T_{1}r),\label{eq:region1_potential}
\end{equation}

\begin{equation}
f_{\phi,3}(r)=c_{2}K_{0}(\tau r)+c_{3}I_{0}(\tau r),
\end{equation}

\begin{equation}
f_{\phi,2}(r)=c_{4}J_{0}(T_{2}r)+c_{5}Y_{0}(T_{2}r),
\end{equation}

\begin{equation}
f_{\phi,4}(r)=c_{6}K_{0}(\tau r)+c_{7}I_{0}(\tau r),
\end{equation}
where $c_{n}$ ($n=1,\ldots,7$ ) are arbitrary constants that are
determined by imposing the boundary conditions in Sec. \ref{subsec:Boundary-Conditions}.
The parameters $T_{1}$ and $T_{2}$ in the arguments of the above
Bessel functions are related to regions 1 and 2, i.e., in the electron
streams, and the parameter $\tau$ is related to the vacuum regions
outside of the electron streams (regions 3 and 4), defined as

\begin{equation}
T_{1}^{2}=\tau^{2}\left(\dfrac{\left(\beta_{\mathrm{p,1}}\right)^{2}-\left(k-\beta_{0,1}\right)^{2}}{\left(k-\beta_{0,1}\right)^{2}}\right),
\end{equation}

\begin{equation}
T_{2}^{2}=\tau^{2}\left(\dfrac{\left(\beta_{\mathrm{p,2}}\right)^{2}-\left(k-\beta_{0,2}\right)^{2}}{\left(k-\beta_{0,2}\right)^{2}}\right),
\end{equation}

\begin{equation}
\tau^{2}=\left(k^{2}-k_{0}^{2}\right).\label{eq:Gamma-1-1-1}
\end{equation}

In these equations, $\beta_{\mathrm{p,1}}=\omega_{\mathrm{p,1}}/u_{0,1}$
and $\beta_{\mathrm{p,2}}=\omega_{\mathrm{p,2}}/u_{0,2}$ are the
plasma phase constants related to stream 1 and stream 2, respectively,
related to the two plasma frequencies $\omega_{\mathrm{p,1}}=\sqrt{\eta\rho_{0,1}/\varepsilon_{0}}$
and $\omega_{\mathrm{p,2}}=\sqrt{\eta\rho_{0,2}/\varepsilon_{0}}$.
Note that we did not consider the Bessel's function of the second
kind (Neumann's function) $Y_{0}(T_{1}r)$ in $f_{\phi,1}(r)$ because
the scalar potential should be finite at $r=0$, and $Y_{0}(T_{1}r)$
has a singularity at $r=0$. When the six boundary conditions in Eq.
(\ref{eq:Contin}) are enforced, together with the PEC condition at
the tunnel wall in Eq. (\ref{eq:Wall}), the resulting set of seven
equations are cast in matrix form as $\mathbf{\underline{M}c}=\mathbf{0}$,
where $\mathbf{c}=\left[\begin{array}{ccccccc}
c_{1}, & c_{2}, & c_{3}, & c_{4}, & c_{5}, & c_{6}, & c_{7}\end{array}\right]^{\mathrm{T}}$ ($\mathrm{T}$ denotes the transpose operation) and the matrix $\underline{\mathbf{M}}$
is defined as

\newpage
\begin{strip}
\rule{\dimexpr(0.5\textwidth-0.5\columnsep-0.4pt)}{\wttlinewidth}
\par
\parindent \parindentlength
\begin{scriptsize}
\begin{equation}
\mathbf{\underline{M}}=\left[\begin{array}{ccccccc}
J_{0}\left(T_{1}R_{1}\right) & \mathrm{-}K_{0}\left(\tau R_{1}\right) & -I_{0}\left(\tau R_{1}\right) & 0 & 0 & 0 & 0\\
T_{1}J_{0}^{\prime}\left(T_{1}R_{1}\right) & -\tau K_{0}^{\prime}\left(\tau R_{1}\right) & -\tau I_{0}^{\prime}\left(\tau R_{1}\right) & 0 & 0 & 0 & 0\\
0 & K_{0}\left(\tau R_{\mathrm{i},2}\right) & I_{0}\left(\tau R_{\mathrm{i,2}}\right) & -J_{0}\left(T_{2}R_{\mathrm{i,2}}\right) & -Y_{0}\left(T_{2}R_{\mathrm{i,2}}\right) & 0 & 0\\
0 & \tau K_{0}^{\prime}\left(\tau R_{\mathrm{i},2}\right) & \tau I_{0}^{\prime}\left(\tau R_{\mathrm{i},2}\right) & -T_{2}J_{0}^{\prime}\left(T_{2}R_{\mathrm{i},2}\right) & -T_{2}Y_{0}^{\prime}\left(T_{2}R_{\mathrm{i},2}\right) & 0 & 0\\
0 & 0 & 0 & J_{0}\left(T_{2}R_{\mathrm{o,2}}\right) & Y_{0}\left(T_{2}R_{\mathrm{o,2}}\right) & -K_{0}\left(\tau R_{\mathrm{o,2}}\right) & -I_{0}\left(\tau R_{\mathrm{o,2}}\right)\\
0 & 0 & 0 & T_{2}J_{0}^{\prime}\left(T_{2}R_{\mathrm{o,2}}\right) & T_{2}Y_{0}^{\prime}\left(T_{2}R_{\mathrm{o,2}}\right) & -\tau K_{0}^{\prime}\left(\tau R_{\mathrm{o,2}}\right) & -\tau I_{0}^{\prime}\left(\tau R_{\mathrm{o,2}}\right)\\
0 & 0 & 0 & 0 & 0 & K_{0}\left(\tau R_{\mathrm{t}}\right) & I_{0}\left(\tau R_{\mathrm{t}}\right)
\end{array}\right].\label{eq:BigM}
\end{equation}

\end{scriptsize}
\par
\hfill
\rule[0.5\baselineskip]{\dimexpr(0.5\textwidth-0.5\columnsep-1pt)}{\wttlinewidth}
\end{strip} 

A solution exists when one finds $k$ such that $\mathrm{det}(\mathbf{\underline{M}})=0.$
Here, we look for complex $k$ solutions, though they may also be
purely real. The matrix $\mathbf{\underline{M}}$ may become ill-conditioned
(i.e., the condition number of the matrix becomes large \cite{cline1979estimate})
when using an imaginary part of the wavenumber $k$ that makes the
Bessel functions extremely large or small in value. This also occurs
when $k$ is nearly equal to $\beta_{0,1}$ or $\beta_{0,2}$, which
are the poles of $T_{1}$ and $T_{2}$, respectively. To overcome
this issue, we follow the same procedure as in \cite{ramo1939space},
i.e., we reduce the number of equations describing the boundary conditions
until we obtain only one characteristic equation, $C_{e}$, that is
described in terms of the space charge wavenumber $k$ of the system.
Numerical solutions for $k$ which make $\left|C_{e}\right|=0$ are
eigenmode solutions of the two-stream system.

\section{Illustrative Examples}

As an illustrative example, we consider an electron beam consisting
of two streams that have equivalent kinetic dc voltages of $V_{0,1}=7\ \mathrm{kV}$
and $V_{0,2}=6\ \mathrm{kV}$, corresponding to average electron speeds
of $u_{0,1}=0.164c$ and $u_{0,2}=0.152c$, respectively, from the
relativistic relation $V_{0}=\left(\sqrt{1-(u_{0}/c)^{2}}-1\right)c^{2}/\eta$.
Stream 1 is a solid cylinder with a circular cross-section and has
an outer radius $R_{1}=0.1\ \mathrm{mm}$. Stream 2 is annular in
cross section, with inner and outer radii of $R_{\mathrm{i},2}=1\ \mathrm{mm}$
and $R_{\mathrm{o},2}=1.1\ \mathrm{mm}$, respectively. The metallic
tunnel is made of a PEC and has an inner radius of $R_{t}=2\ \mathrm{mm}$,
as illustrated in Fig. \ref{fig:Schematic}. The dc currents of stream
1 and stream 2 are $I_{0,1}=A_{1}\rho_{0,1}u_{0,1}=50\ \mathrm{mA}$
and $I_{0,2}=A_{2}\rho_{0,2}u_{0,2}=50\ \mathrm{mA}$, respectively,
corresponding to dc volume charge densities of $\rho_{0,1}=0.0325\ \mathrm{C}/\mathrm{m}^{3}$
and $\rho_{0,2}=0.00166\ \mathrm{C}/\mathrm{m}^{3}$ for streams 1
and 2, respectively.

We show in Fig. \ref{Fig:CE_vs_Complexk} the magnitude of the characteristic
equation, $C_{e}$, defined in Appendix \ref{sec:CE-def-and-mode-profile},
in log scale when the real and imaginary parts of $k$ are swept at
a fixed frequency of $f=5\ \mathrm{GHz}$. The roots of the characteristic
equation correspond to locations in Fig. \ref{Fig:CE_vs_Complexk}
where $C_{e}$ tends to zero (shown by the dark blue regions). We
label the four modes in Fig. \ref{Fig:CE_vs_Complexk} that correspond
to the dominant modes of the system. We also verify that, for the
four $k$ solutions, the scalar potential function $f_{\phi}(r)$
is continuous at the radii, where we have the boundaries between the
electron stream and vacuum regions. The points in the complex $k$
plane of Fig. \ref{Fig:CE_vs_Complexk} where $C_{e}$ is not vanishing
(i.e., everywhere, except for the labeled four solutions) correspond
to scalar potential functions $f_{\phi}(r)$ that have discontinuities
at the radii corresponding to the boundaries between regions. Thus,
these points are not valid solutions. The radial profiles of the scalar
electric potential for the four dominant modes labeled in Fig. \ref{Fig:CE_vs_Complexk}
are shown in Fig. \ref{Fig:case1_Mode1}, Fig. \ref{Fig:case1_Mode2},
Fig. \ref{Fig:case1_Mode3} and Fig. \ref{Fig:case1_Mode4}.

\begin{figure}[tbh]
\begin{centering}
\centering \subfigure[]{\includegraphics[width=0.99\columnwidth]{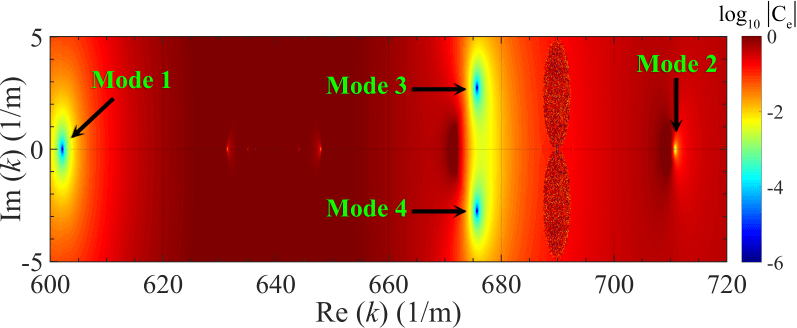}\label{Fig:CE_vs_Complexk}}
\par\end{centering}
\begin{centering}
\subfigure[]{\includegraphics[width=0.48\columnwidth]{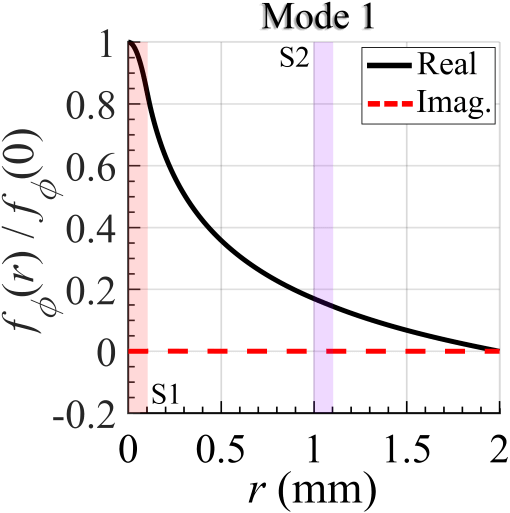}\label{Fig:case1_Mode1}}\subfigure[]{\includegraphics[width=0.48\columnwidth]{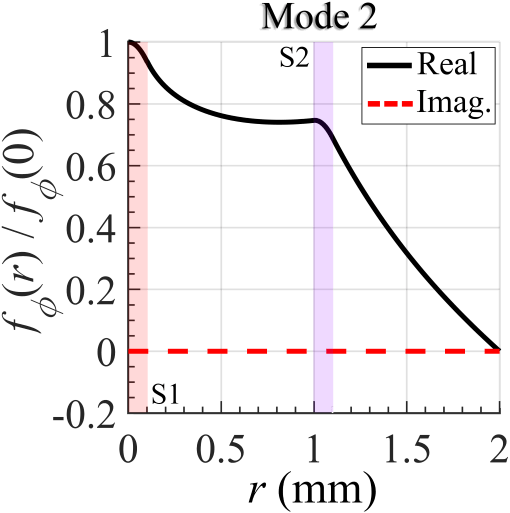}\label{Fig:case1_Mode2}}
\par\end{centering}
\begin{centering}
\subfigure[]{\includegraphics[width=0.48\columnwidth]{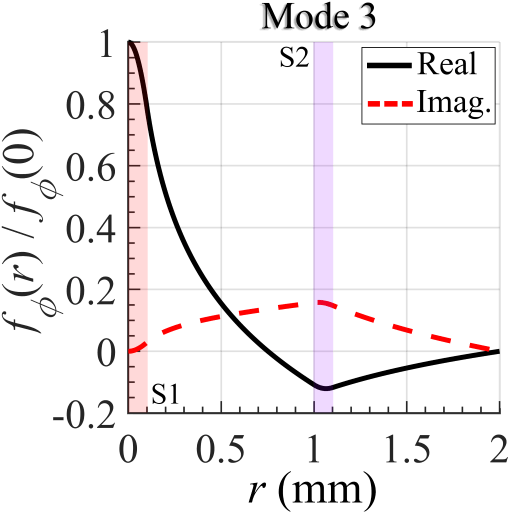}\label{Fig:case1_Mode3}}\subfigure[]{\includegraphics[width=0.48\columnwidth]{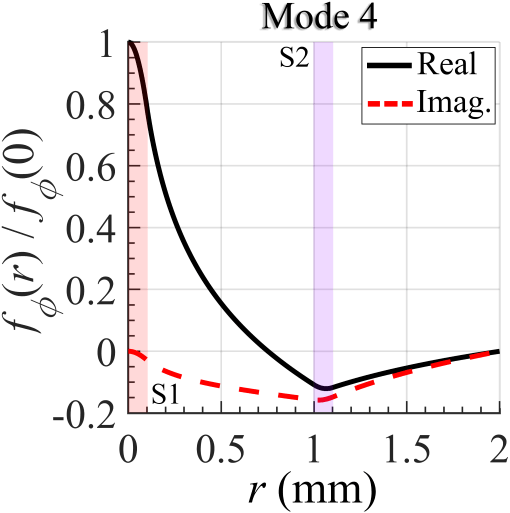}\label{Fig:case1_Mode4}}
\par\end{centering}
\centering{}\caption{(a) Values of $C_{e}$ versus complex wavenumber $k$, evaluated at
$f=5\ \mathrm{GHz}$. The labeled points where $\mathrm{C_{e}\to0}$
represent the wavenumber solutions of the system. (b)-(e) Normalized
potential profiles corresponding to modes labeled in (a). The radial
distribution of the potential function confirms the validity of the
solution because one can observe the continuity of the potential function
and its derivative at the boundaries between the electron streams
and vacuum and its vanishing at the tunnel wall at $R_{t}=2\ \mathrm{mm}$.
The shaded regions in (b)-(e) represent the radial locations of stream
1 and stream 2.\label{Fig:case1All}}
\end{figure}

\begin{figure}[tbh]
\begin{centering}
\includegraphics[width=0.9\columnwidth]{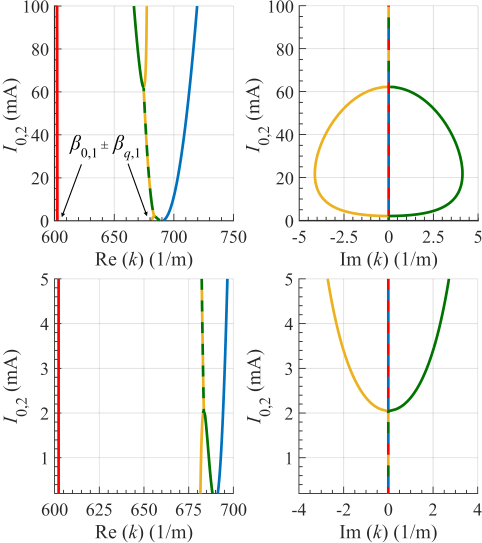}
\par\end{centering}
\centering{}\caption{Top: Modal dispersion diagram of the four complex-valued space-charge
wavenumbers as a function of stream 2 dc current. Bottom: magnified
version around the EPD for lower values of the current $I_{0,2}$.\label{Fig:dispersion_I02_swept}}
\end{figure}

In Fig. \ref{Fig:dispersion_I02_swept}, we show the four modes of
the two-stream system when the dc current of stream 2 is swept, while
the dc current of stream 1 is held constant at $I_{0,1}=50\ \mathrm{mA}$.
The frequency is fixed at $f=5\ \mathrm{GHz}$, and the equivalent
kinetic stream voltages are held constant at $V_{0,1}=7\ \mathrm{kV}$
and $V_{0,2}=6\ \mathrm{kV}$. Also, the study can be used to verify
the validity of the mode-finding method that we use. When the dc current
of stream 2 approaches zero, this is equivalent to a case where stream
1 only exists in the tunnel. Therefore, when $\mathrm{I_{0,2}\to0}$
one sees only two solutions that coincide with the two conventional
plasma modes that have wavenumbers described as $k=\beta_{0,1}\pm\beta_{q,1}$,
where $\beta_{q,1}=R_{sc}\omega_{p,1}/u_{0,1}$, and $R_{sc}$ is
the plasma frequency reduction factor, calculated using the method
shown in \cite{branch1955plasma}. Figure \ref{Fig:dispersion_I02_swept}
shows that there exist two transition points (bifurcations) close
to $I_{0,2}=2\ \mathrm{mA}$ and $I_{0,2}=62.2\ \mathrm{mA}$, between
which, exponentially growing space-charge waves occur due to the two-stream
instability effect, which happens when there is a sufficient velocity
difference between electron streams and sufficient stream currents,
as predicted in \cite{pierce1949double,pierce1949new} using an abstract
theoretical model. The bifurcation points in Fig. \ref{Fig:dispersion_I02_swept}
are exceptional points of degeneracy (EPDs) \cite{hanson2018exceptional,rouhi2021exceptional,figotin2021exceptional},
which are conditions where two or more eigenmodes coalesce in their
wavenumbers and eigenvectors. In Appendix \ref{sec:Swapping-Stream-Velocities}
we show the effect of swapping the two stream velocities on the resulting
space charge wavenumbers. We find that the dispersion diagram does
not change significantly compared to the case studied in this section.

\begin{figure}[tbh]
\begin{centering}
\includegraphics[width=0.7\columnwidth]{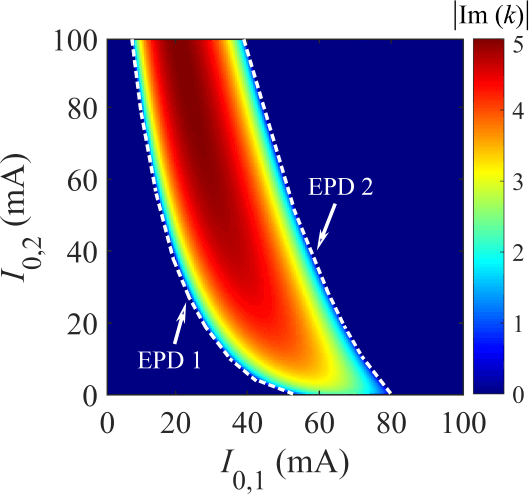}
\par\end{centering}
\centering{}\caption{Contour showing transition boundaries in dashed white curves (labeled
EPD 1 and EPD 2), where the imaginary part of the space-charge wavenumber
vanishes. Between these boundaries, $\mathrm{Im}(k)\protect\neq0$
and two-stream instability occurs for different combinations of stream
dc currents.\label{Fig:Contour_Current}}
\end{figure}

To understand the conditions resulting in the beam instability, we
have swept both dc currents of stream 1 and stream 2 at a fixed frequency
of $f=5\ \mathrm{GHz}$ and fixed equivalent kinetic stream voltages
of $V_{0,1}=7\ \mathrm{kV}$ and $V_{0,2}=6\ \mathrm{kV}$, and we
monitored the imaginary part of the wavenumber of the growing mode
to observe what conditions result in instability. We show in Fig.
\ref{Fig:Contour_Current} the absolute value of the imaginary part
of the wavenumber for the modes that exhibit instability, which is
depicted by the color area between the two dashed white curves. We
see that $\mathrm{Im}(k)=0$ outside of the dashed white curves (i.e.,
shown by dark blue). The dashed white boundary between the region
where the wavenumber has $\mathrm{Im}(k)=0$ and the colored region
where $\mathrm{Im}(k)\neq0$ is a curve showing the transition points,
or EPDs, where the two-stream instability begins to occur as labeled
in Fig. \ref{Fig:Contour_Current}. Fixing $I_{0,1}=50\ \mathrm{mA}$
and sweeping $I_{0,2}$ in Fig. \ref{Fig:Contour_Current}, results
in a curve very similar to that shown in Fig. \ref{Fig:dispersion_I02_swept}
depicting the imaginary part of the wavenumber showing the two bifurcation
points at $I_{0,2}=2\ \mathrm{mA}$ and $I_{0,2}=62.2\ \mathrm{mA}$.
Figure \ref{Fig:Contour_Current} indicates that charge wave amplification
occurs only for certain combinations of stream 1 and stream 2 currents
for a given set of frequency and beam parameters.

\begin{figure}[tbh]
\begin{centering}
\includegraphics[width=0.7\columnwidth]{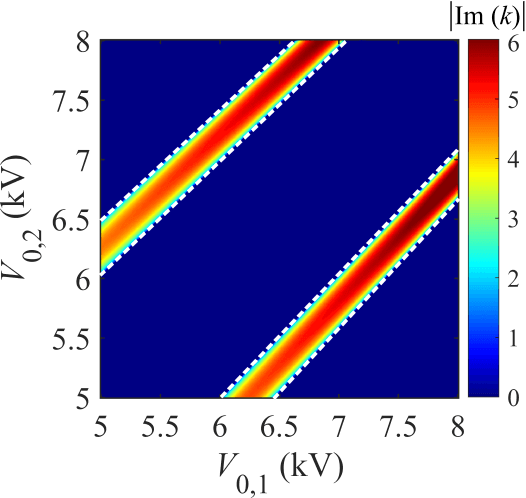}
\par\end{centering}
\centering{}\caption{Contour showing the transition boundaries where the imaginary part
of the space-charge wavenumber vanishes. The colored region between
dashed white lines represents the region where $\mathrm{Im}(k)\protect\neq0$
and two-stream instability occurs for different combinations of stream
dc voltages. The dashed white lines indicate the transition boundaries
that separate the regions where $\mathrm{Im}(k)=0$ from regions where
$\mathrm{Im}(k)\protect\neq0$ that lead to stream instability.\label{Fig:Contour_Voltage}}
\end{figure}

We repeat the previous study when both dc voltages of stream 1 and
stream 2 are swept at a fixed frequency of $f=5\ \mathrm{GHz}$ and
the beam currents are held constant at $I_{0,1}=50\ \mathrm{mA}$
and $I_{0,2}=50\ \mathrm{mA}$. We show in Fig. \ref{Fig:Contour_Voltage}
the absolute value of the imaginary part of the wavenumber for the
mode that exhibits a growing instability. Note that instability occurs
when either stream 1 has a higher dc voltage than stream 2 or vice
versa. Like in the previous figure, the dashed white lines represent
the boundary between the stability and instability regions. For the
studied range of stream 1 and stream 2 dc voltages in Fig. \ref{Fig:Contour_Voltage},
one finds that instability occurs when the difference between the
equivalent kinetic dc voltages of the two streams is approximately
$1\ \mathrm{kV}\lessapprox|V_{0,1}-V_{0,2}|\lessapprox1.5\ \mathrm{kV}$.

\begin{figure}[tbh]
\centering{}\includegraphics[width=0.9\columnwidth]{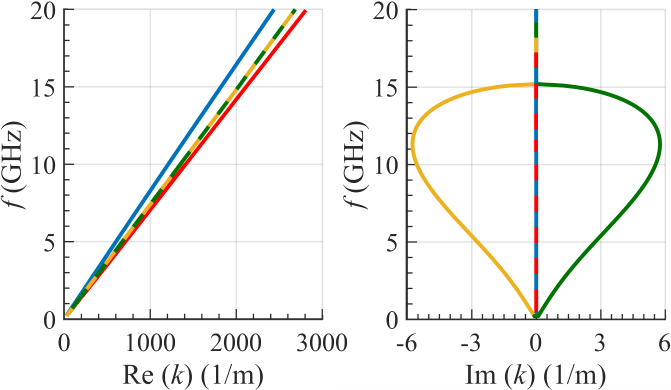}\caption{Modal dispersion showing the four complex space-charge wavenumbers
as a function of frequency.\label{Fig:Disp_vs_Freq}}
\end{figure}

We show the modal wavenumber-frequency dispersion relation for the
two-stream system in Fig. \ref{Fig:Disp_vs_Freq} when the operating
frequency is swept while the two beam currents are held constant at
$I_{0,1}=50\ \mathrm{mA}$ and $I_{0,2}=50\ \mathrm{mA}$ and the
equivalent stream voltages are held constant at $V_{0,1}=7\ \mathrm{kV}$
and $V_{0,2}=6\ \mathrm{kV}$ (as in the first and second examples
above). The figure shows that the amplification resulting from the
two-stream instability occurs from dc up to a threshold frequency,
which is 15 GHz in this case (note that the cutoff frequency of the
lowest mode ($\mathrm{TE}_{11}$) in a metallic circular waveguide
of radius $R_{t}=2\ \mathrm{mm}$ is approximately 44 GHz). For the
case of a two-stream instability amplifier, which has been experimentally
investigated in works such as \cite{hollenberg1949experimental,haeff1949electron},
it may be potentially beneficial to have a growing instability up
to a threshold frequency that is below the lowest cutoff frequency
of a circular waveguide, since the device will be less susceptible
to regenerative oscillations or backward-wave oscillations that would
otherwise exist on a slow-wave structure in a conventional TWT amplifier,
as explained in \cite{chen1996efficiency}. However, in \cite{phillips1990review},
it was stated that backward-wave oscillations may still be an issue
if long slow-wave structures are used to extract amplified waves from
a two-stream amplifier.

\begin{figure}[tbh]
\begin{centering}
\includegraphics[width=0.9\columnwidth]{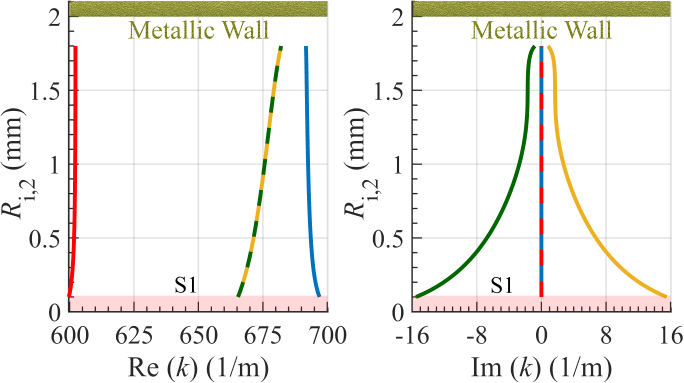}
\par\end{centering}
\centering{}\caption{The real and imaginary parts of the wavenumber $k$ of the four space-charge
modes as a function of the inner radius of stream 2. The shaded region
$\mathrm{S1}$ at the bottom indicates the region where stream 1 exists.\label{Fig:Disp_vs_R}}
\end{figure}

Next, observing the modal dispersion of the wavenumbers as the distance
between the two streams varies, Fig. \ref{Fig:Disp_vs_R}, gives us
a better understanding of how the two streams are coupled. The coupling
is controlled in this example by sweeping the inner radius $R_{\mathrm{i},2}$
of stream 2 while keeping its width, $R_{\mathrm{o},2}-R_{\mathrm{i},2}=0.1\ \mathrm{mm}$,
constant. We also keep the radius of steam 1 constant at $R_{1}=0.1\ \mathrm{mm}$.
Therefore, what varies is the distance between the two streams, $R_{\mathrm{i},2}-R_{1}$.
In Fig. \ref{Fig:Disp_vs_R} we show the four complex wavenumbers
of the two charge-wave eigenmodes versus $R_{\mathrm{i},2}$. The
operating frequency is still kept at $f=5\ \mathrm{GHz}$, the currents
of the streams are $I_{0,1}=50\ \mathrm{mA}$ and $I_{0,1}=50\ \mathrm{mA}$,
and the equivalent kinetic voltages of the streams are still kept
at $V_{0,1}=7\ \mathrm{kV}$ and $V_{0,2}=6\ \mathrm{kV}$. Note that
the radius $R_{\mathrm{i},2}=1\ \mathrm{mm}$ was considered in all
the previous examples, and that for the given values of $I_{0,1}$,
$I_{0,2}$, $V_{0,1}$, and $V_{0,2}$ considered here, the two-stream
electron beam exhibits instability, as was shown in Fig. \ref{Fig:dispersion_I02_swept}.
The plots in Fig. \ref{Fig:Disp_vs_R} reveal that when the distance
between the two beams $R_{\mathrm{i},2}-R_{1}$ gets smaller, the
$\mathrm{Im}(k)$ of the unstable eigenmode gets larger. Conversely,
when the distance between the streams gets larger, $\mathrm{Im}(k)$
gets smaller. Furthermore, for all values of the distance $R_{i,2}-R_{1}$,
there are always two eigenmodes with purely real $k$, represented
by the red and blue curves, and the difference between their $k$
values remains more or less constant for all considered distances
$R_{i,2}-R_{1}$.

\section{Conclusion}

We have provided an analytical method to determine the wavenumbers
and associated electric fields of space-charge waves supported by
an electron beam made of two coaxial streams. Our analytical model
has determined the two-stream instability regions and the bifurcation
points in the wavenumber-dispersion diagrams for a coaxial two-stream
system. Specific combinations of dc current and dc equivalent kinetic
voltages cause instability. Additionally, this instability growth
rate is enhanced when the two streams are closer together. The findings
of this work can also be used to build an equivalent transmission
line model that describes the two streams' interaction, in analogy
to what was done earlier in the analysis of single-stream TWTs and
BWOs \cite{pierce1947theoryTWT,pierce1951waves,figotin2013multi,tamma2014extension,rouhi2021exceptional};
such a model can be used in the analysis of amplifiers based on two
interacting streams, as proposed in \cite{figotin2021analytic,evans2019instability}.
The method presented in this paper is useful for modeling electron
beam systems consisting of two coaxial streams, such as those generated
by multi-stream electron guns described in \cite{neben2021co,islam2022modeling,islam2022multiple}.

\appendices{}

\section{General Solution of the Scalar Electric Potential Function in a Region
with Moving Electrons (Stream Region)\label{sec:General-Solution-of-phi-stream}}

We provide the basic steps to find the general solution that satisfies
the differential equation governing the electron beam and the electromagnetic
fields. For a given region containing a single electron stream (i.e.,
either region 1 or 2 in Fig. \ref{Fig:Front_View}), the electrons
in that stream are assumed to be traveling only in the axial direction
with uniform dc velocity $u_{0}$ and uniform dc charge density $\rho_{0}$
(e.g., region 1 in Fig. \ref{Fig:Front_View} has $u_{0}=u_{0,1}$
and $\rho_{0}=\rho_{0,1})$. We assume that the electron beam and
the electromagnetic dynamics follow a time-harmonic wave function
$e^{j\omega t-jkz}$ that involves modulation in the charges' axial
speed and charge density, which in cylindrical coordinates are written
as

\begin{equation}
\begin{array}{c}
u(r,z,t)=u_{0}+\Re\left(u_{\mathrm{m}}(r)e^{j\omega t-jkz}\right),\end{array}
\end{equation}

\begin{equation}
\begin{array}{c}
\rho(r,z,t)=\rho_{0}+\Re\left(\rho_{\mathrm{m}}(r)e^{j\omega t-jkz}\right),\end{array}
\end{equation}

\begin{equation}
\begin{array}{c}
\phi(r,z,t)=\Re\left(f_{\phi}(r)e^{j\omega t-jkz}\right),\end{array}
\end{equation}

\begin{equation}
\begin{array}{c}
\mathbf{A}(r,z,t)=\Re\left(f_{A}(r)e^{j\omega t-jkz}\right)\ \hat{\mathbf{z}}.\end{array}
\end{equation}

Because of the strong external axial magnetic field that confines
the beam, we have assumed that the beam modulation occurs only along
the axial direction. Starting from the divergence relation of the
magnetic vector potential $\nabla\cdot\mathbf{A}=-\mu_{0}\varepsilon_{0}\partial\phi/\partial t$,
one finds that

\begin{equation}
f_{A}(r)=\dfrac{\omega\mu_{0}\varepsilon_{0}}{k}f_{\phi}(r).\label{eq:A_to_Phi_simplified}
\end{equation}

The axial electric field component is found using the relation $\mathbf{E}=-\nabla\phi-\partial\mathbf{A}/\partial t$,
which yields

\begin{equation}
\begin{array}{c}
E_{z}(r,z,t)=-\dfrac{\partial\phi}{\partial z}-\dfrac{\partial\left(\mathbf{A\cdot}\hat{\mathbf{z}}\right)}{\partial t}\\
\begin{array}{ccccc}
 &  &  &  & =\Re\left(\left(jkf_{\phi}(r)-j\omega f_{A}(r)\right)e^{j\omega t-jkz}\right).\end{array}
\end{array}\label{eq:Ez_expr1}
\end{equation}

We use the relation in Eq. (\ref{eq:A_to_Phi_simplified}) to simplify
Eq. (\ref{eq:Ez_expr1}) as

\begin{equation}
E_{z}(r,z,t)=\Re\left(\left(j\dfrac{k^{2}-\omega^{2}\mu_{0}\varepsilon_{0}}{k}\right)f_{\phi}(r)e^{j\omega t-jkz}\right).\label{eq:Ez_expr1-1}
\end{equation}

Newton's second law that describes the equation of motion for electrons
(for a strongly confined beam of electrons, i.e., with no radial or
azimuthal motion) is written as $m\thinspace du/dt=-eE_{z}$. First,
we express the total derivative of the velocity of the electrons in
the phasor domain as

\begin{equation}
\begin{array}{c}
\dfrac{du}{dt}=\dfrac{\partial u}{\partial t}+u_{0}\dfrac{\partial u}{\partial z}\\
\begin{array}{ccccc}
 &  &  &  & =\Re\left(\left(j\omega-jku_{0}\right)u_{\mathrm{m}}(r)e^{j\omega t-jkz}\right).\end{array}
\end{array}
\end{equation}

When this is inserted into Newton's second law applied to the electrons,
we find the relation between the velocity and electric potential functions
as

\begin{equation}
\begin{array}{c}
u_{\mathrm{m}}(r)=\dfrac{\eta}{u_{0}}\dfrac{k_{0}^{2}-k^{2}}{k\left(k-\beta_{0}\right)}f_{\phi}(r),\end{array}\label{eq:eq_motions_S1-1}
\end{equation}
where $k_{0}=\omega\sqrt{\mu_{0}\epsilon_{0}}$ is the free space
wavenumber and $\beta_{0}=\omega/u_{0}$ for the stream-containing
region of interest. Moreover, we consider the continuity equation
or conservation of charge,

\begin{equation}
\dfrac{\partial\left(\rho u\right)}{\partial z}=-\dfrac{\partial\rho}{\partial t},
\end{equation}
which is simplified to

\begin{equation}
\rho_{\mathrm{m}}(r)=\dfrac{k\rho_{0}}{u_{0}\left(\beta_{0}-k\right)}u_{\mathrm{m}}(r).\label{eq:Charge_u_relation}
\end{equation}

By substituting Eq. (\ref{eq:eq_motions_S1-1}) into Eq. (\ref{eq:Charge_u_relation}),
the latter equation yields the relation between the charge density
and the potential function as

\begin{equation}
\rho_{\mathrm{m}}(r)=-\dfrac{\eta\rho_{0}}{u_{0}^{2}}\dfrac{k_{0}^{2}-k^{2}}{\left(k-\beta_{0}\right)^{2}}f_{\phi}(r).\label{eq:Charge_u_relation-1}
\end{equation}

The final equation that is used to find the potential function is
obtained from Gauss' law, $\nabla\cdot\mathbf{E}=\rho/\epsilon_{0}$,
where $\mathbf{E}=-\nabla\phi-d\mathbf{A}/dt$, leading to

\begin{equation}
\nabla^{2}\phi-\frac{\partial}{\partial t}\left(\nabla\cdot\mathbf{A}\right)=-\dfrac{\rho}{\varepsilon_{0}}.\label{eq:Phi_A_DE}
\end{equation}

Then, using the Lorentz gauge $\nabla\cdot\mathbf{A}=-\mu_{0}\varepsilon_{0}\partial\phi/\partial t$,
we obtain the inhomogeneous wave equation that governs the scalar
electric potential $\left(\nabla^{2}-\mu_{0}\varepsilon_{0}\frac{d^{2}}{dt^{2}}\right)\phi=-\rho/\varepsilon_{0}$,
where the charge density is cast in terms of the scalar potential
function $f_{\phi}(r)$ using Eq. (\ref{eq:Charge_u_relation-1}).
Due to the azimuthal symmetry of our system in cylindrical coordinates,
we have $\nabla^{2}\phi=\frac{1}{r}\frac{\partial}{\partial r}\left(r\frac{\partial\phi}{\partial r}\right)+\frac{\partial^{2}\phi}{\partial z^{2}}$.
After taking the time and $z$ derivatives, the final equation to
be solved is

\begin{equation}
\dfrac{d^{2}f_{\phi}(r)}{dr^{2}}+\dfrac{1}{r}\dfrac{df_{\phi}(r)}{dr}+T^{2}f_{\phi}(r)=0,\label{eq:D.E_Bessel}
\end{equation}
where

\begin{equation}
T^{2}=\left(k^{2}-k_{0}^{2}\right)\left(\dfrac{\beta_{\mathrm{p}}^{2}-\left(k-\beta_{0}\right)^{2}}{\left(k-\beta_{0}\right)^{2}}\right).\label{eq:Gamma}
\end{equation}

Here, $\beta_{p}=\omega_{p}/u_{0}$, and $\omega_{p}=\sqrt{\eta\rho_{0}/\varepsilon_{0}}$
is the plasma frequency of the electron stream in the region considered
(either stream 1 or stream 2). The general solution of the differential
equation in Eq. (\ref{eq:D.E_Bessel}) is written in terms of Bessel's
functions as

\begin{equation}
f_{\phi}(r)=a_{1}J_{0}(Tr)+a_{2}Y_{0}(Tr),\label{eq:general_soln_stream_JY}
\end{equation}
where $a_{1}$ and $a_{2}$ are arbitrary constants, $J_{0}$ is the
Bessel function of the first kind and order zero, and $Y_{0}$ is
the Bessel function of the second kind (Neumann's function) of order
zero.

One may also find the general solution in terms of modified Bessel's
functions, as follows. By transforming the differential equation in
Eq. (\ref{eq:D.E_Bessel}) using $\kappa=jr$, as was done in \cite{mclachlan1961bessel},
we find

\begin{equation}
\dfrac{d^{2}g(\kappa)}{d\kappa^{2}}+\dfrac{1}{\kappa}\dfrac{dg(\kappa)}{d\kappa}-T^{2}g(\kappa)=0,\label{eq:D.E_Bessel-2}
\end{equation}
which has a general solution $g(\kappa)=d_{1}I_{0}(T\kappa)+d_{2}K_{0}(T\kappa).$
Thus, one may rewrite the general solution of Eq. (\ref{eq:general_soln_stream_JY})
in terms of modified Bessel functions as $f_{\phi}(r)=d_{1}I_{0}(jTr)+d_{2}K_{0}(jTr).$

\section{General Solution of the Scalar Electric Potential Function in an
Empty Region (Vacuum)\label{sec:General-Solution-of-phi-vacuum}}

We consider the case where the studied region is empty, i.e., it does
not contain charges. Starting from Eq. (\ref{eq:Phi_DE}), the steps
are exactly as the previous case in Sec. \ref{sec:General-Solution-of-phi-stream},
except that in the vacuum region $\rho(r)=0$. This leads to the homogeneous
(i.e., source-free) wave equation that governs the scalar electric
potential 

\begin{equation}
\dfrac{d^{2}f_{\phi}(r)}{dr^{2}}+\dfrac{1}{r}\dfrac{df_{\phi}(r)}{dr}-\tau^{2}f_{\phi}(r)=0,\label{eq:D.E_Bessel-1}
\end{equation}
where

\begin{equation}
\tau^{2}=k^{2}-k_{0}^{2}.\label{eq:Gamma-1-1}
\end{equation}

The general solution of the differential equation in Eq. (\ref{eq:D.E_Bessel-1})
is a linear combination of modified Bessel functions, 

\begin{equation}
f_{\phi}(r)=b_{1}I_{0}(\tau r)+b_{2}K_{0}(\tau r),
\end{equation}
where $b_{1}$ and $b_{2}$ are arbitrary constants, $I_{0}$ is the
modified Bessel function of the first kind and order zero, and $K_{0}$
is the modified Bessel function of the second kind and order zero.

\section{Characteristic Equation Definition and Mode Profile\label{sec:CE-def-and-mode-profile}}

We show the steps we used to find solutions to the system of equations
in $\mathbf{\underline{M}c}=\mathbf{0}$, where $\mathbf{c}=\left[\begin{array}{ccccccc}
c_{1}, & c_{2}, & c_{3} & c_{4}, & c_{5}, & c_{6}, & c_{7}\end{array}\right]^{\mathrm{T}}$ and the matrix $\mathbf{\underline{M}}$ is given in Eq. (\ref{eq:BigM}).
Once we assume the stream parameters and the radii of the structure,
the only unknowns we are left with are the 7 constants in $\mathbf{c}$
and the wavenumber $k$. First, we assume that the potential function
at the center of stream region 1 (given by Eq. (\ref{eq:region1_potential}))
is normalized such that $f_{\phi,1}(r=0)=c_{1}=1\:\mathrm{V}$.

Then, for a given wavenumber $k$, one finds the rest of the constants
of the system by solving the last six equations described in $\mathbf{\underline{M}c}=\mathbf{0}$,
which yields

\begin{strip}
\rule{\dimexpr(0.5\textwidth-0.5\columnsep-0.4pt)}{\wttlinewidth}
\par
\parindent \parindentlength
\begin{scriptsize}

\begin{equation}
\left[\begin{array}{c}
c_{2}\\
c_{3}\\
c_{4}\\
c_{5}\\
c_{6}\\
c_{7}
\end{array}\right]=\left[\begin{array}{cccccc}
\tau K_{0}^{\prime}\left(\tau R_{1}\right) & \tau I_{0}^{\prime}\left(\tau R_{1}\right) & 0 & 0 & 0 & 0\\
K_{0}\left(\tau R_{\mathrm{i},2}\right) & I_{0}\left(\tau R_{\mathrm{i},2}\right) & -J_{0}\left(T_{2}R_{\mathrm{i},2}\right) & -Y_{0}\left(T_{2}R_{\mathrm{i},2}\right) & 0 & 0\\
\tau K_{0}^{\prime}\left(\tau R_{\mathrm{i},2}\right) & \tau I_{0}^{\prime}\left(\tau R_{\mathrm{i},2}\right) & -T_{2}J_{0}^{\prime}\left(T_{2}R_{\mathrm{i},2}\right) & -T_{2}Y_{0}^{\prime}\left(T_{2}R_{\mathrm{i},2}\right) & 0 & 0\\
0 & 0 & J_{0}\left(T_{2}R_{\mathrm{o,2}}\right) & Y_{0}\left(T_{2}R_{\mathrm{o,2}}\right) & -K_{0}\left(\tau R_{\mathrm{o,2}}\right) & -I_{0}\left(\tau R_{\mathrm{o,2}}\right)\\
0 & 0 & T_{2}J_{0}^{\prime}\left(T_{2}R_{\mathrm{o,2}}\right) & T_{2}Y_{0}^{\prime}\left(T_{2}R_{\mathrm{o,2}}\right) & -\tau K_{0}^{\prime}\left(\tau R_{\mathrm{o,2}}\right) & -\tau I_{0}^{\prime}\left(\tau R_{\mathrm{o,2}}\right)\\
0 & 0 & 0 & 0 & K_{0}\left(\tau R_{\mathrm{t}}\right) & I_{0}\left(\tau R_{\mathrm{t}}\right)
\end{array}\right]^{-1}\left[\begin{array}{c}
T_{1}J_{0}^{\prime}\left(T_{1}R_{1}\right)\\
0\\
0\\
0\\
0\\
0
\end{array}\right].\label{eq:Mode}
\end{equation}

\end{scriptsize}
\par
\hfill
\rule[0.5\baselineskip]{\dimexpr(0.5\textwidth-0.5\columnsep-1pt)}{\wttlinewidth}
\end{strip}

Reaching this stage, we found the seven constants in $\mathbf{c}$,
which satisfy six out of the seven boundary conditions. The next step
is that we solve to find the last unknown $k$ that will satisfy the
first boundary condition as

\begin{equation}
C_{e}=J_{0}\left(T_{1}R_{1}\right)-c_{2}K_{0}\left(\tau R_{1}\right)-c_{3}I_{0}\left(\tau R_{1}\right)=0.\label{CE}
\end{equation}

We find solutions of $k$ by searching for complex wavenumbers that
guarantee that Eq. (\ref{CE}) is satisfied, which implicitly guarantees
that the remaining equations are also satisfied since $c_{2}$ and
$c_{3}$ are found based on satisfying the rest of the boundary conditions.
The modes' profile in Fig. \ref{Fig:CE_vs_Complexk} is found based
the constants calculated from Eq. (\ref{eq:Mode}) with $c_{1}=1\:\mathrm{V}$
to have a normalized electric potential at the center of stream 1
as $f_{\phi}(r=0)=1\:\mathrm{V}$.

\section{Effect of Swapping Stream Velocities\label{sec:Swapping-Stream-Velocities}}

\begin{figure}[h]
\begin{centering}
\centering \subfigure[]{\includegraphics[width=0.99\columnwidth]{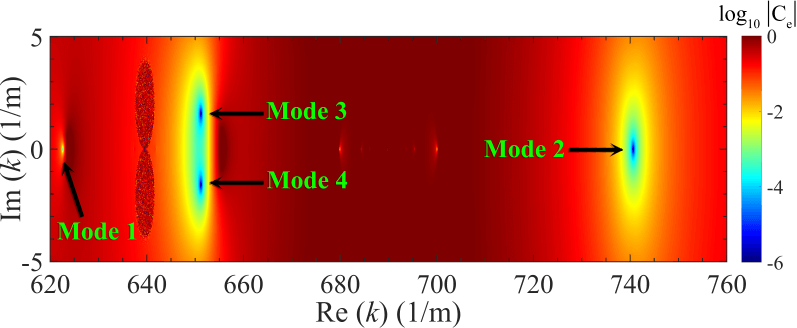}\label{Fig:CE_vs_Complexk-1}}
\par\end{centering}
\begin{centering}
\subfigure[]{\includegraphics[width=0.48\columnwidth]{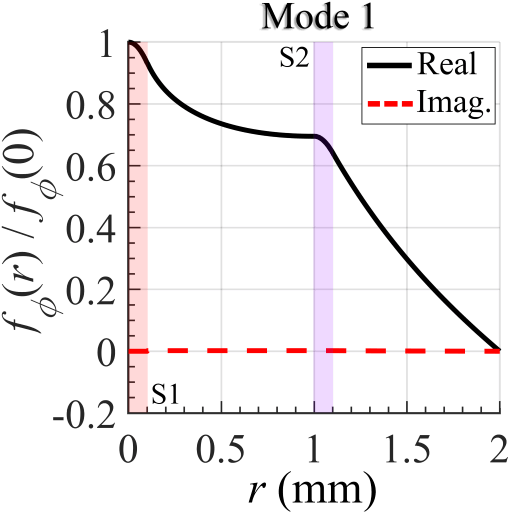}\label{Fig:case1_Mode1-1}}\subfigure[]{\includegraphics[width=0.48\columnwidth]{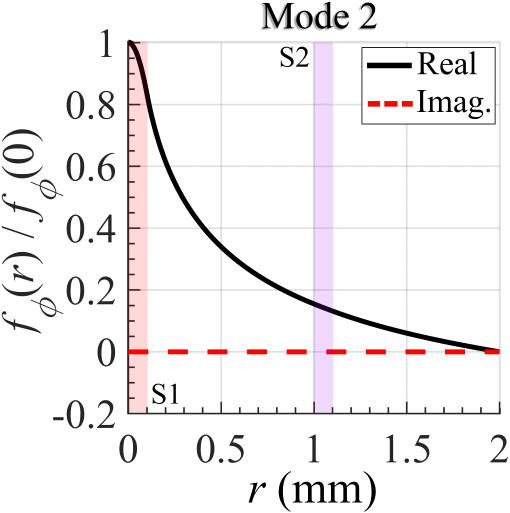}\label{Fig:case1_Mode2-1}}
\par\end{centering}
\begin{centering}
\subfigure[]{\includegraphics[width=0.48\columnwidth]{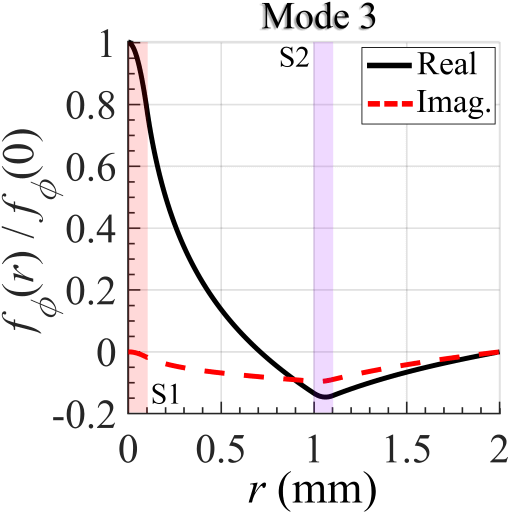}\label{Fig:case1_Mode3-1}}\subfigure[]{\includegraphics[width=0.48\columnwidth]{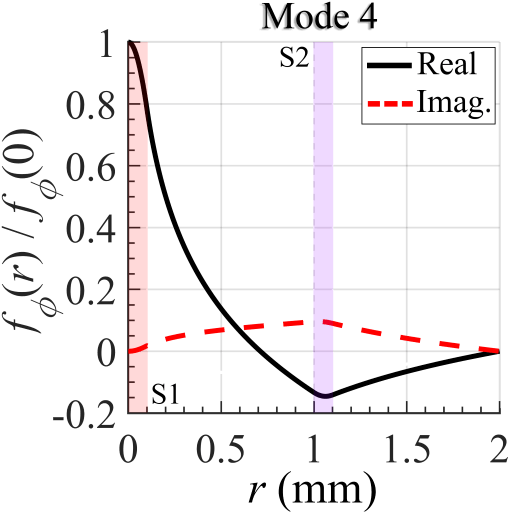}\label{Fig:case1_Mode4-1}}
\par\end{centering}
\centering{}\caption{(a) Values of $C_{e}$ versus complex wavenumber $k$, evaluated at
$f=5\ \mathrm{GHz}$, when the two streams dc voltages are $V_{0,1}=6\ \mathrm{kV}$
and $V_{0,2}=7\ \mathrm{kV}$. (b)-(e) Normalized potential profiles
corresponding to the four modes labeled in (a). These plots should
be compared with those in Fig. \ref{Fig:case1All}, with the difference
that the stream voltages have been inverted since in Fig. \ref{Fig:case1All}
we had $V_{0,1}=7\ \mathrm{kV}$ and $V_{0,2}=6\ \mathrm{kV}$.}
\end{figure}

We show in Fig. \ref{Fig:CE_vs_Complexk-1} the magnitude of the value
of the characteristic equation in log scale for the same case with
results shown in Fig. \ref{Fig:CE_vs_Complexk}, except that the beam
dc voltages of the two streams are swapped, i.e., $V_{0,1}=6\ \mathrm{kV}$
and $V_{0,2}=7\ \mathrm{kV}$ instead of $V_{0,1}=7\ \mathrm{kV}$
and $V_{0,2}=6\ \mathrm{kV}$. We label the four modes in Fig. \ref{Fig:CE_vs_Complexk-1}
that correspond to the dominant modes of the system. Compared to the
previous case, we still see that the interaction between the two streams
results in instability.

\begin{figure}[h]
\begin{centering}
\includegraphics[width=0.9\columnwidth]{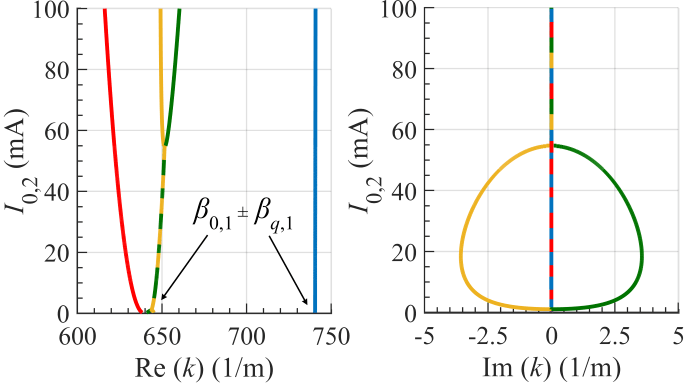}
\par\end{centering}
\centering{}\caption{Modal dispersion diagram of the complex-valued space-charge wavenumber
as a function of stream 2 dc current. In this case we have $V_{0,1}=6\ \mathrm{kV}$
and $V_{0,2}=7\ \mathrm{kV}$. These results should be compared with
those in Fig. \ref{Fig:dispersion_I02_swept}, with the difference
that the stream voltages have been here inverted since in Fig. \ref{Fig:dispersion_I02_swept}
we had $V_{0,1}=7\ \mathrm{kV}$ and $V_{0,2}=6\ \mathrm{kV}$.\label{fig:dispersion_I02_swept-1}}
\end{figure}

In Fig. \ref{fig:dispersion_I02_swept-1}, we show the four modes
of the two-stream system when the dc current of stream 2 is swept
similarly to case shown in Fig. \ref{Fig:dispersion_I02_swept}, except
that the beam dc voltages of the two streams are swapped, i.e., now
$V_{0,1}=6\ \mathrm{kV}$ and $V_{0,2}=7\ \mathrm{kV}$, instead of
$V_{0,1}=7\ \mathrm{kV}$ and $V_{0,2}=6\ \mathrm{kV}$ as in Fig.
\ref{Fig:dispersion_I02_swept}. The figures show that there exist
two transition points (bifurcations) close to $I_{0,2}=1\ \mathrm{mA}$
and $I_{0,2}=54.7\ \mathrm{mA}$, between which, exponentially growing
space-charge waves occur due to the two-stream instability effect.
This is similar to the other case in Fig. \ref{Fig:dispersion_I02_swept},
except that the transition points in Fig. \ref{Fig:dispersion_I02_swept}
occurred at approximately $I_{0,2}=2\ \mathrm{mA}$ and $I_{0,2}=62.2\ \mathrm{mA}$.

\newpage\bibliographystyle{ieeetr}

\end{document}